\documentstyle[osa,manuscript]{revtex}  

\begin{document}

\title{Dilation of the Giant Vortex State in a Mesoscopic Superconducting Loop. }
\author{S. Pedersen\dag, G.R. Kofod, J.C. Hollingbery, C.B. S\o rensen, and P.E. Lindelof}
\address{The Niels Bohr Institute, University of Copenhagen, Universitetsparken 5, DK-2100 Copenhagen, Denmark}
\date{\today}
\maketitle
\begin{abstract}
We have experimentally investigated the magnetisation of a mesoscopic aluminum loop at temperatures well below
the superconducting transition temperature $T_{c}$. The flux quantisation of the superconducting loop was
investigated with a $\mu$-Hall magnetometer in magnetic field intensities between $\pm 100 {\rm Gauss}$. The magnetic field
intensity periodicity observed in the magnetization measurements is expected to take integer values of the 
superconducting flux quanta $\Phi_{0}=h/2e$. A closer inspection of the periodicity, however, reveal a 
sub flux quantum shift. This fine structure we interpret as a consequence of a so called giant vortex 
state nucleating towards either the inner or the outer side of the loop. These findings are in agreement with recent theoretical reports.   
\end{abstract}
Ever since the original observation and explanation of flux quantization \cite{Deaver,Doll}, the superconducting flux quanta 
$\Phi_{0}=h/2e$ has played a fundamental role in solid state physics.
The concept of flux quantisation has been crucial for the interpretation of a wide range of classical condensed matter 
experiments, concerning e.g. weakly connected rings \cite{Silver,Jackel1,Jackel2} 
and Little-Parks oscillations \cite{Little,Parks,Groff}.\newline
However, all these investigations were primarily performed at temperatures close to the critical temperature $T_{c}$ and at magnetic field intensities well below $H_{c2}$.
Recently it has become possible with $\mu$-Hall magnetometers to perform high resolution magnetisation experiments on
small superconducting aluminium discs in the full magnetic field intensity range of superconductivity and at temperatures well 
below $T_{c}$ \cite{Geim1,Geim2,Geim3}. These investigations have revealed information from deep within the superconducting phase, a regime which 
previously hasn't been accessible. Not unexpectedly these reports have attracted considerable interest also from a theoretical point of view \cite{Deo,Deo2,Schweigert2,Schweigert,Palacios,benoist,peeters3,peeters4,vital}.\newline
It is well known that for type-II $(\kappa=\lambda/\xi>1/\sqrt{2})$ bulk superconductors a triangular Abrikosov vortex lattice is created in the magnetic field intensity range $H_{c1}<H<H_{c2}$ where $\kappa$
is the Ginzburg-Landau parameter, and $H_{c1}$ and $H_{c2}$ are the first and second critical fields. 
Since the effective Ginzburg-Landau parameter is significantly increased in thin films when the width of the film becomes comparable to the superconducting coherence length $\xi_{o}$, the appearance of
an Abrikosov lattice is expected even in thin films consisting of type-I superconducting materials. 
When the spatial dimensions of the sample are decreased even further, and several length scales of the system 
become comparable with $\xi_{o}$, the competition between the Abrikosov vortex configuration and symmetry of the 
sample boundary becomes important. Hence for such mesoscopic systems the bulk critical fields $H_{c1}$ and $H_{c2}$
no longer are the only controlling parameters of the vortex configurations.\newline
When considering sufficiently small superconducting rings the confinement effects from the boundaries are
dominating and impose a circular symmetry on the superconducting order parameter. Hence the order parameter is
expected to be given by $\psi(r)=F(r)e^{i L \theta}$ where $L$ is the angular momentum or vorticity of the vortex. 
When the superconductor is described by such a circular symmetric order parameter it is said to be in a {\em giant vortex state}\cite{peeters3,peeters4,vital}.\newline
In a recent theoretical work the properties of giant vortex states and multi-vortex states in mesoscopic superconducting discs and rings were treated extensively\cite{peeters3,peeters4}. It was found that the giant vortex state indeed is energetic
favorable in narrow rings due to the strong influence of the ring surface. Furthermore, the superconducting state can consist of a combination of the paramagnetic- and the diamagnetic Meissner state. In other words the direction of the supercurrents closest to the outer edge are opposite to the currents running closest to the
inner edge. This means that at a certain effective radius between the outer and inner edge the supercurrent density
goes to zero. Since this effective zero-current radius is the one that determines the area in which the flux is quantized, it becomes
possible to measure this effective radius by studying the magnetization of superconducting mesoscopic loops.   
It was furthermore pointed out that when increasing the magnetic field intensity from zero field this effective radius would move
towards the outer edge as a signature of the giant vortex state.\newline
The measurement described in this paper was performed on a micron sized superconducting aluminium loop placed 
on top of a $\mu$-Hall magnetometer. The $\mu$-Hall magnetometer was etched out of a  $\rm GaAs/Ga_{0.7}Al_{0.3}As$ 
heterostructure. The mobility and electron density of the bare two-dimensional electron gas was $\mu=42 \rm{T^{-1}}$ and $n=1.9\times 10^{15} {\rm m^{-2}}$. A symmetrical $4 {\rm \mu m}\times 4 {\rm \mu m}$ Hall geometry was defined by standard e-beam lithography on top of the heterostructure. In a later processing step a lift-off mask was defined on top of the $\mu$-Hall probe by e-beam lithography. After deposition of a $t=90 {\rm nm}$ thick layer of aluminium and lift-off the sample looked as presented in Fig.\ref{fig1}.
The mean radius of the aluminium loop is $R=2.16 {\rm \mu m}$ and the average wire width $w$ is $316\pm40 {\rm nm}$. The superconducting coherence length was estimated to be approximately $\xi_{o}= 180{\rm nm}$ corresponding to a bulk critical field of $H_{c2}=\Phi_{0}/2\pi\xi^{2}_{o}\approx 100 {\rm Gauss}$. \newline
By using the expression
\begin{center}
\begin{equation}
n\Phi_{0}=n\frac{h}{2e}=\Delta(\mu_{0}H)\pi R^{2},
\label{eq}
\end{equation} 
\end{center}
where $A=\pi R^{2}$ is the area of the loop given by its mean radius $R$, it is found that a single flux jumps ($n=1$) corresponds to a magnetic field periodicity given by $\Delta(\mu_{0}H)=1.412 {\rm Gauss}$ for the ring shown in Fig.\ref{fig1}.\newline
The samples was cooled in a $\rm{^{3}He}$ cryostat equipped with a superconducting soleniode driven by a DC 
current supply. The magnetic field intensity was changed in steps of $57.7 {\rm mGauss}$. Measurements discussed here 
were performed in the temperature range between $T=0.3{\rm K}$ and the transition temperature of the 
superconducting loop $T_{c} \approx 1.2{\rm K}$.\newline 
The relation between the Hall voltage $V_{H}$ and the magnetic field intensity $H$ perpendicular to the $\mu$-Hall magnetometer is given by the classical Hall effect
\begin{equation}
V_{H}=-\frac{I}{ne}\mu_{0}(H+\alpha M),
\end{equation}
were $I$ is the DC current through the $\mu$-Hall magnetometer and $\alpha$ is a dimensionless number of the 
order of unity, which corresponds to the ratio between the sensitive area of the $\mu$-Hall probe and the area of 
the object which is the source of the magnetisation $M$ \cite{peeters1,peeters2}. For our superconducting rings we find that $\alpha$ typically was in the range between $0.3 \dots 0.4$.\newline
By using standard AC lock-in techniques, where the driving current $I$ was modulated, the Hall voltage $V_{H}$ 
was measured as a function of magnetic field intensity $\mu_{0}H$.\newline
Similar results as the ones presented here were observed in several samples with identical dimensions in a number of cooldowns.\newline
Also a circular loops with a width of $w=630{\rm nm}$, but with the same mean radius as the loops described 
above, were investigated. \newline
In Fig.\ref{mag}. is displayed the measured local magnetsation $\mu_{0}M$ detected by the $\mu$-Hall probe 
as a function of magnetic field intensity $\mu_{0}H$. The measurement was performed at $T=0.36{\rm K}$ on the device presented 
in Fig.\ref{fig1}. The curve displays a series of distinct jumps corresponding to the abrupt changes in magnetisation of the superconducting loop. The difference in magnetic field intensity between two successive flux jumps is approximately given by $\Delta(\mu_{0}H)=1.4{\rm Gauss}$ or $\Delta(\mu_{0}H)=2.8{\rm Gauss}$ which corresponds to either single or double flux jumps ($n=1$ or $n=2$).\newline
Large flux jumps $(n>1)$ or flux avalanches, occur whenever the system is trapped in a metastable state. 
It was generally observed that these flux avalanches become more pronounced with decreasing temperature, 
at low magnetic field intensities and for wide loops. Furthermore the flux avalanches were sensitive 
to the cooling procedure.\newline 
The energy barrier causing the metastability of the eigenstates of the loop, are due to either the 
Beam-Livingston surface barrier or the volume barrier, or even an interplay of 
both \cite{Deo2,bean,price}.\newline
In Fig.\ref{flux}. the magnetic field intensity difference between successive jumps $\Delta(\mu_{0}H)$, in units of 
the $1.412 {\rm Gauss}$ (corresponding to a single superconducting flux quantum), have been plotted as a function 
of magnetic field intensity. It is seen that the magnetic field intensity difference between the observed jumps is, to a 
high accuracy, given by integer values of $1.412 {\rm Gauss}$. At absolute magnetic field intensities lower than $40 
{\rm Gauss}$ double flux jumps dominate, whereas at higher absolute magnetic field intensities only single flux jumps 
are observed. The figure shows both an up-sweep and a down-sweep as indicated by the arrows. \newline
Similar results obtained from the device with width $w=630{\rm nm}$ are presented in Fig.\ref{tykke}. For these 
thicker loops it is seen that the flux avalanches are much more pronounced; avalanches corresponding to eleven 
single flux jumps were observed around zero magnetic field intensity. For these loops a gradual transition from huge 
flux avalanches ($n=11$) to single flux jumps occur as the magnetic field intensity is increased - similar to 
the sharp transition between double and single flux jumps observed for the thinner loops.\newline
In the graphs presented in Fig.\ref{flux}. it is seen that a small systematic variation of the value of the 
flux jumps occur when the magnetic field intensity is changed. This fine structure appears as a memory effect, in the 
sense that as the magnetic field intensity is increased (decreased) the size of the flux jumps decreases (increases). 
Thus these deviations are dependent, not only on the size of the magnetic field intensity, but also on which direction 
the magnetic field intensity was sweept during measurements. 
The data presented in Fig.\ref{flux}. has been replotted on Fig.\ref{radius}. in the following way: We use 
Eq.(\ref{eq}) to calculate the effective radius $R$ of the superconducting loop and plot this radius as a 
function of magnetic field intensity. The dotted horizontal lines in Fig.\ref{radius}. represents the mean inner $R_{i}$ 
and outer radius $R_{o}$ determined from the SEM picture. It is seen that as the magnetic field intensity is changed 
from negative to positive values, the effective radius, as defined from the flux quantization condition of the 
loop, changes from inner to outer radius and vice versa. \newline
For a superconducting loop at low magnetic field intensities, it is expected that the appropriate effective radius is 
given by the geometrical mean value of the outer and inner radius $R=\sqrt{R_{i}R_{o}}$\cite{zharkov,peeters3,peeters4}. This is indeed in good 
agreement with the observed behavior around zero magnetic field intensity.\newline 
In the regime of high magnetic field intensities the concept of surface superconductivity becomes important 
and the giant vortex state will nucleate on the edges of the loop\cite{peeters3,peeters4,vital}. In this regime two degenerate current carrying situations are possible \cite{tinkham} - 
hence the giant vortex state can either circulate the loop clockwise or anti-clockwise.\newline 
Since the orientation of the current in the loop is determined by the sweep direction (Lenz' law), a decreasing 
(increasing) magnetic field intensity will give rise to a anti-clockwise (clockwise) circulation. Hence as the magnetic field 
intensity is sweept from e.g. a large positive value to a large negative value the effective radius of the loop 
will change from inner to outer radius and vice versa giving rise to the observed memory effect.\newline
The width of the giant vortex state is approximately given by the magnetic length 
$l_{H}=\sqrt{\hbar/eH}$\cite{benoist}. Hence any variation of the effective radius should take place over a 
magntic field range given by the condition that the width of the loop and the magnetic length 
are comparable; $w=l_{H}$. Such an estimate gives a characteristic magnetic field intensity of $34\rm{Gauss}$ in good 
agreement with the presented data on Fig.\ref{radius}.\newline    
A similar effective radius analysis of the data presented on Fig.\ref{tykke}. becomes rather dubious due to the 
combination of large flux avalanches and the larger width $w$.\newline 
At even larger magnetic field intensities ($|\mu_{o}H| \approx 60{\rm Gauss}$) the effective radius undergoes a
transition from outer $R_{o}$ (or inner radius $R_{i}$) to the mean radius $R$. We speculate that this could
be due a 2D-1D transition due to an increase in the superconducting coherence length $\xi_{o}$ with magnetic field intensity \cite{vital}.\newline    
The characteristic dimensionless parameter used to distinguish between a discs and loops is given by the ratio $x=R_{i}/R_{o}$ between outer and inner radius \cite{peeters3,peeters4,vital}. In our 
case the thin loops have $x=0.86$, and for the thick loop to $x=0.75$.\newline
In the recent works by two theoretical groups \cite{peeters3,peeters4,vital} it is found that at large $x$ values (corresponding to a loop consisting of a one-dimensional 
wire) no or little variation of the effective radius should be observed. Whereas at small $x$ values 
(corresponding to a disc) a fast decrease of the effective radius occur as the magnetic field intensity increases. In 
the intermediate regime $x=0.5$, a rather smooth transition between average and outer radius should take place 
when the magnetic field intensity increases.\newline
In the presented measurement for the thinner loop ($x=0.86$), we indeed observe that the effective radius varies 
smoothly between inner and outer radius. This behavior looks similar to the one predicted for loops with 
$x=0.5$, however not similar to the one predicted expected for $x=0.75$. We do not find this discrepancy severe for
the following reasons: The calculations by Bruyndoncx et al.\cite{vital} were done using the linearised 
first Ginzburg-Landau equation, hence these results are only valid close to the phase transition, 
viz. $R_{o}/\xi_{o} < 1$. In the work by Peeters et al. \cite{peeters3,peeters4} the full set of non-linear 
Ginzburg-Landau equations were solved self-consistently, in the two cases where $R_{o}/\xi_{o}=4$ and $2$. 
Neither of these conditions were fulfilled in our experiments, where we estimate $R_{o}/\xi_{o}\approx 12$. It is 
furthermore seen by studying the results of Peeters et al. that calculations with larger 
values of $R_{o}/\xi_{0}$ probably would give rise to a better agreement.\newline
For the thick loops ($x=0.75$) we observed large flux avalanches at low magnetic field intensities. The large flux 
avalanches disguise any variation of the effective radius. Furthermore occurrence of flux avalanches in 
superconducting loops have not been dealt with quantitatively in the theoretical literature as far as the 
authors know. Hence comparisons with theory are not possible at the present time.\newline  
In summary, we present high resolution magnetisation measurements performed on superconducting aluminum loops. 
The resolution of the $\mu$-Hall magnetometer allowed us to resolve sub flux quantum effects and hence directly
observe the dilation of a giant vortex state.\newline
This work was financially supported by Velux Fonden, Ib Henriksen Foundation, Novo Nordisk Foundation, The Danish Research Council (grant 9502937, 9601677 and 9800243) and the Danish Technical Research Council (grant 9701490). The authors acknowledge Lars Melwyn Jensen, J. Berger, F. Peeters and V.V. Moshchalkov for discussions.\newline
\dag Present address: Department of Microelectronics and Nanoscience, Chalmers University of Technology, SE 412 96 G\"oteborg, Sweden.

\begin{figure}
\caption[Til LOF]{\small{Scanning electron microscope image of a $\mu$-Hall probe, the cross section of the etched $\mu$-Hall probe is $4 \times 4 \ \mu \rm{m^{2}}$. The mean radius of the superconducting aluminium loop deposited on top of the $\mu$-Hall magnetometer is $2.16 {\rm \mu m}$, and the difference between the outer and inner radius is $314 {\rm nm}$.}\label{fig1} \protect}
\end{figure}

\begin{figure}
\caption[Til LOF]{\small{Measured magnetisation $\mu_{0}M$ detected by the $\mu$-Hall probe as a function of magnetic field intensity $\mu_{0}H$ of the device presented on Fig.(\ref{fig1}). The curve displays distinct jumps corresponding to the abrupt changes in magnetisation of the superconducting loop when the system changes state. The measurements were performed at $T=0.36{\rm K}$.}\protect}
\label{mag}
\end{figure}

\begin{figure}
\caption[Til LOF]{\small{The magnetic field intensity difference $\Delta (\mu_{0}H)$ between two successive jumps in magnetisation. Given in units of $1.412 {\rm Gauss}$ corresponding to a single flux quantum $\Phi_{0}=h/2e$. The plotted jumps are given as a function of magnetic field intensity. The measurement was performed at $T=0.36{\rm K}$. The positive (negative) flux values corresponds to the case where $\mu_{0}H$ was decreased (increased) during the measurements. Arrows indicate sweep direction.} \protect}
\label{flux}
\end{figure}

\begin{figure}
\caption[Til LOF]{\small{Effective radius $R$ calculated by using Eq.\ref{eq}. The data points are the same as the ones presented in Fig.\ref{flux}. Due to the fact that the measurements were performed by stepping the magnetic field intensity with a finite step, the effective radius is only measured with a precision of approximately $40 {\rm nm}$. The filled (open) dots corresponds to single flux jumps $n=1$ (double flux jumps $n=2$). The horizontal lines corresponds to the outer and inner radius determined from the SEM pictures. The arrows indicate sweep direction. It is seen that as the magnetic field intensity is changed the effective radius changes between inner and outer radius of the loop, a changed which depends on sweep direction and magnetic field intensity. The large spread of the data at high magnetic fields corresponds to  regions where the amplitude of the oscillations measured by the Hall probe are small.}\protect}
\label{radius}
\end{figure}

\begin{figure}
\caption[Til LOF]{\small{The magnetic field intensity difference $\Delta (\mu_{0}H)$ between two successive jumps in magnetisation. Given in units of $1.412 {\rm Gauss}$ corresponding to a single flux quantum $\Phi_{0}=h/2e$ for a loop with a width of $w=630{\rm nm}$. The plotted jumps are given as a function of magnetic field intensity. The measurement was performed at $T=0.38{\rm K}$. The positive (negative) flux values corresponds to the case where $\mu_{0}H$ was decreased (increased) during the measurements. Arrows indicate sweep direction.}\protect}
\label{tykke}
\end{figure}

%
%

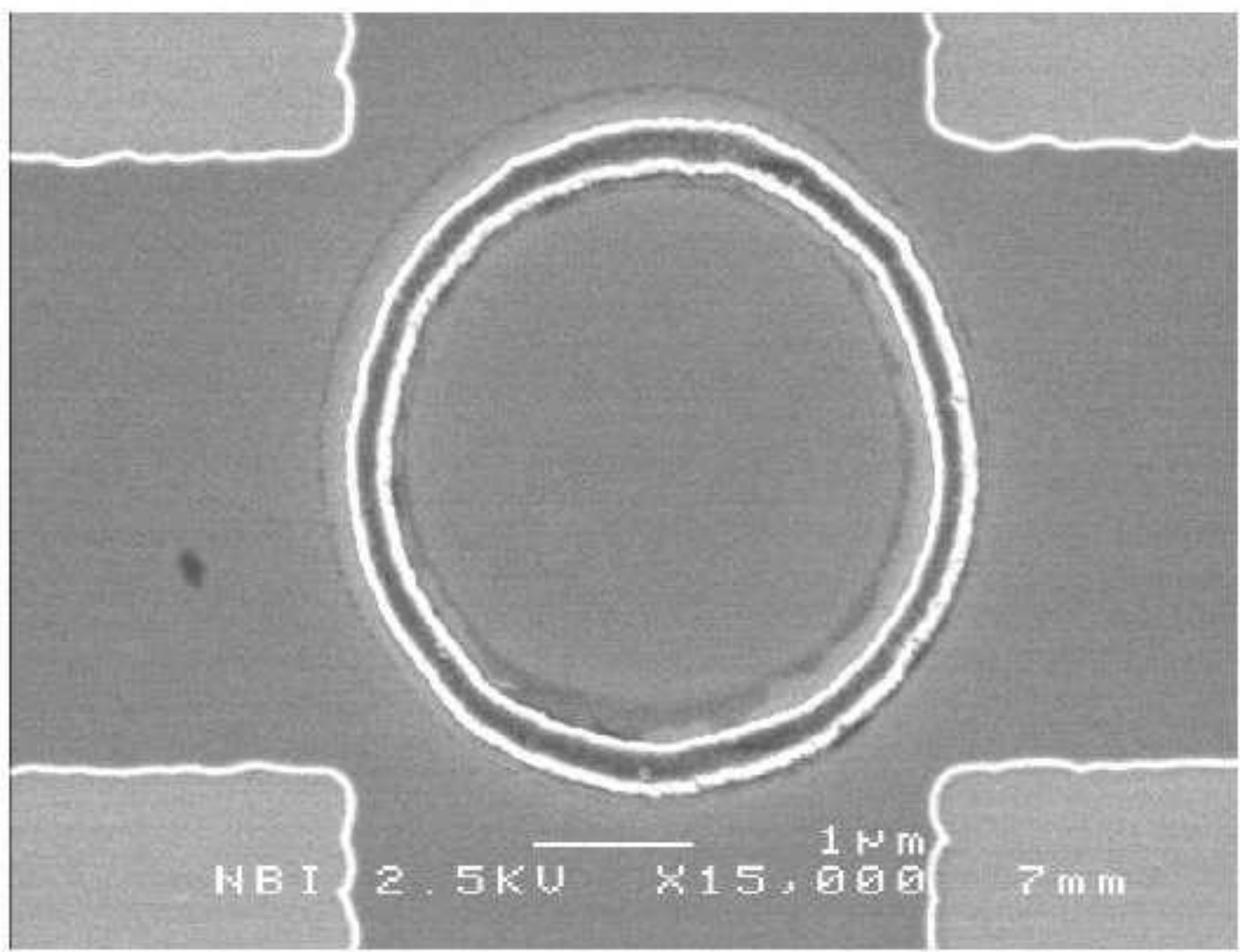
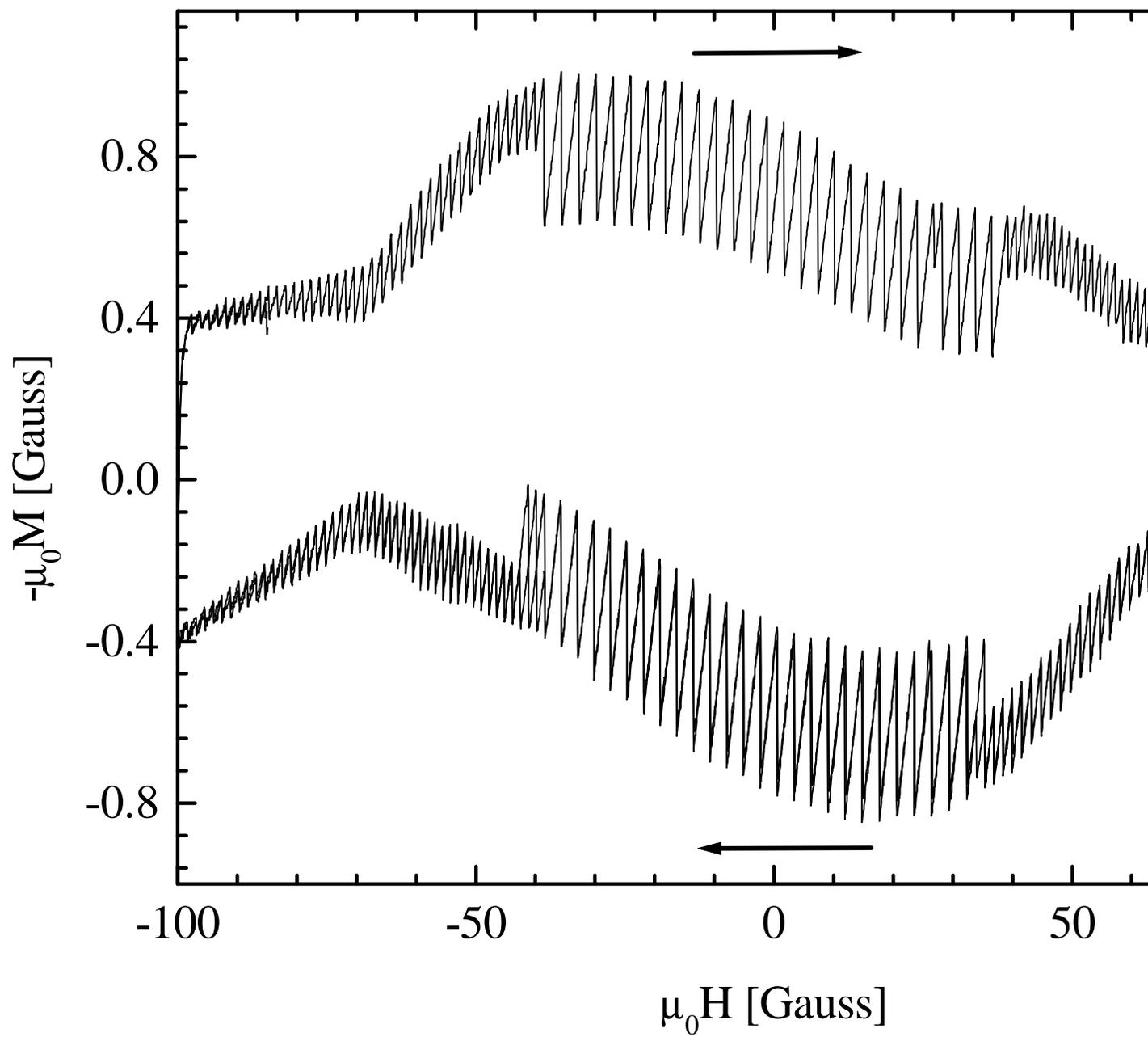
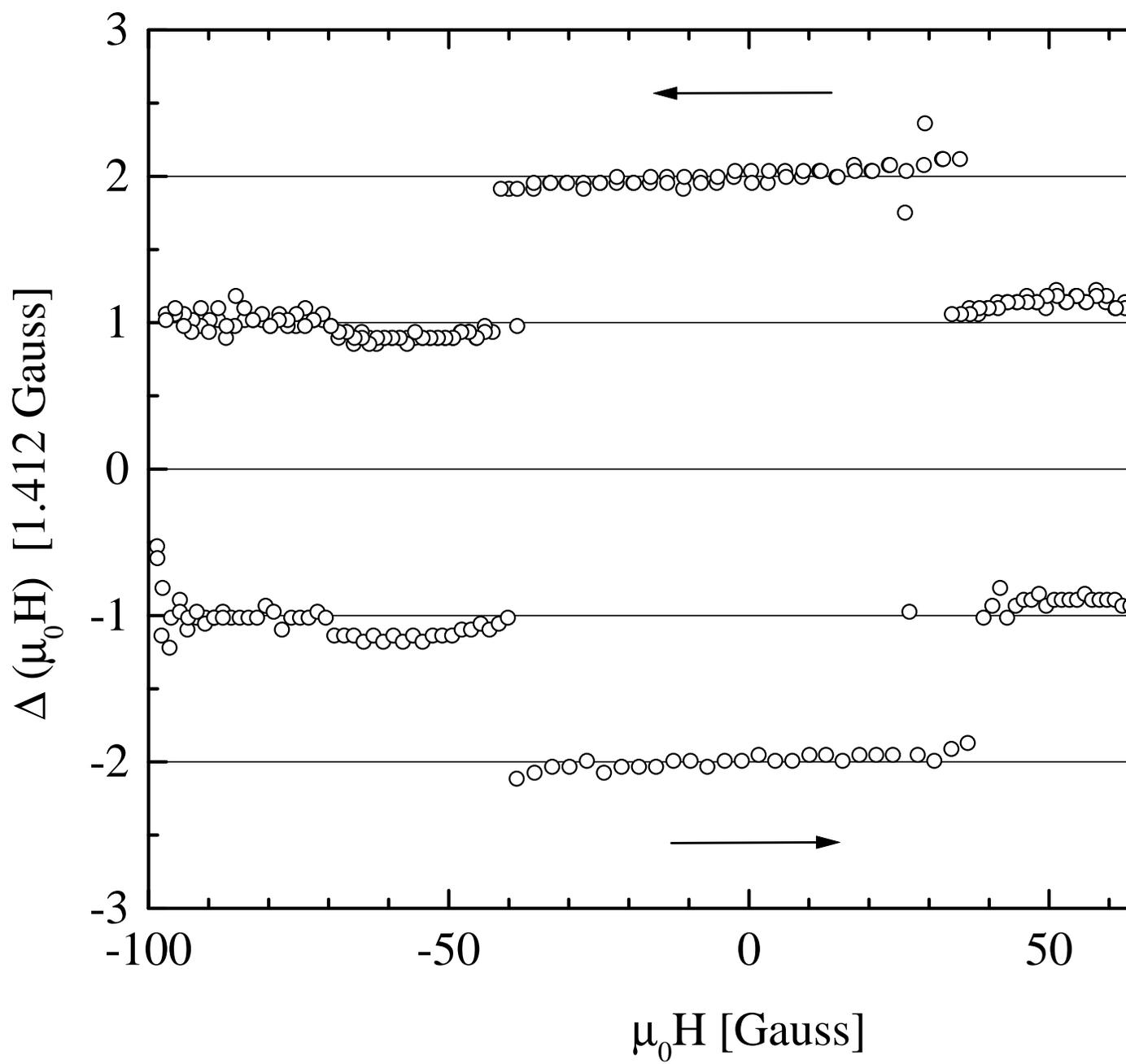
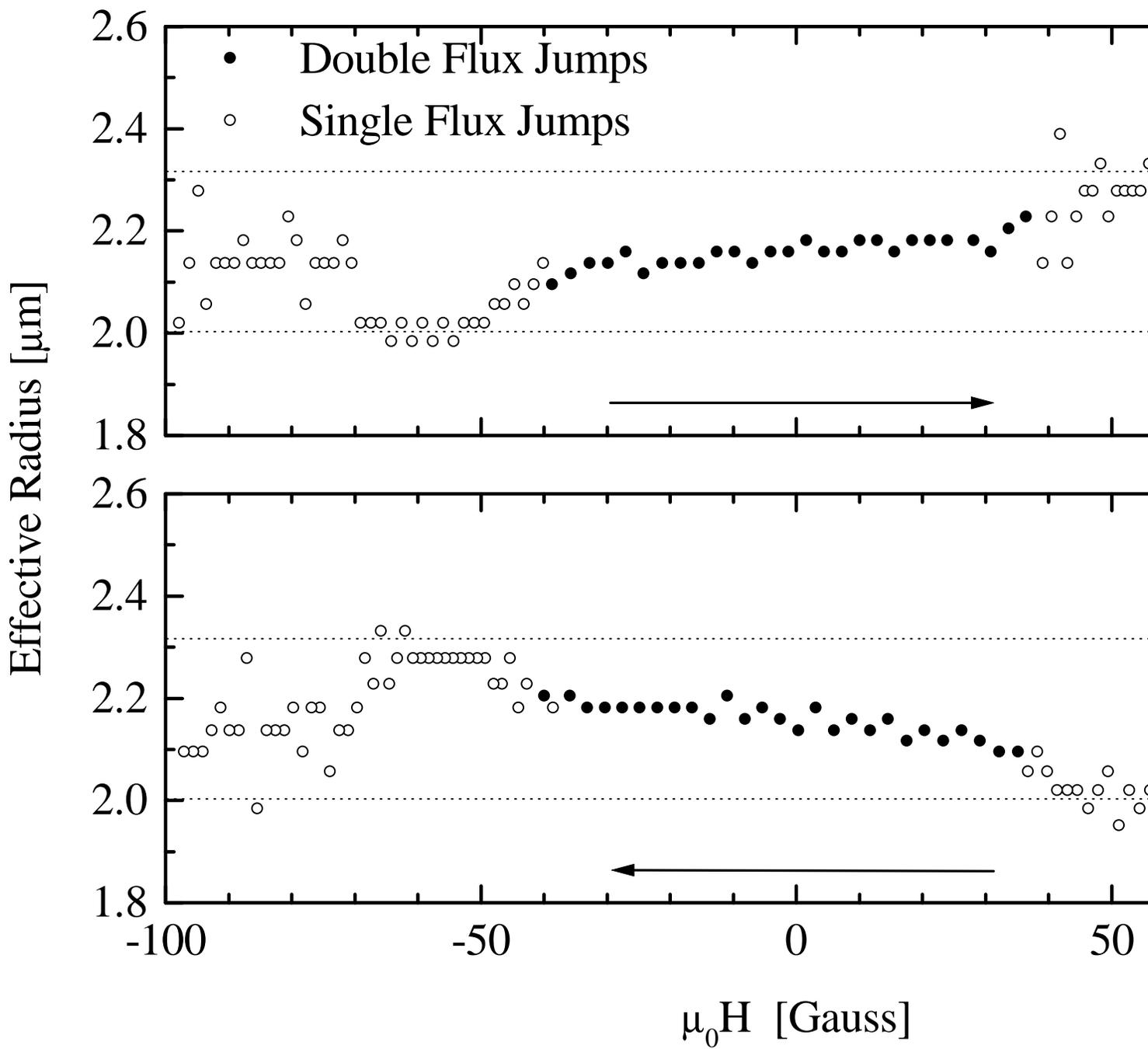
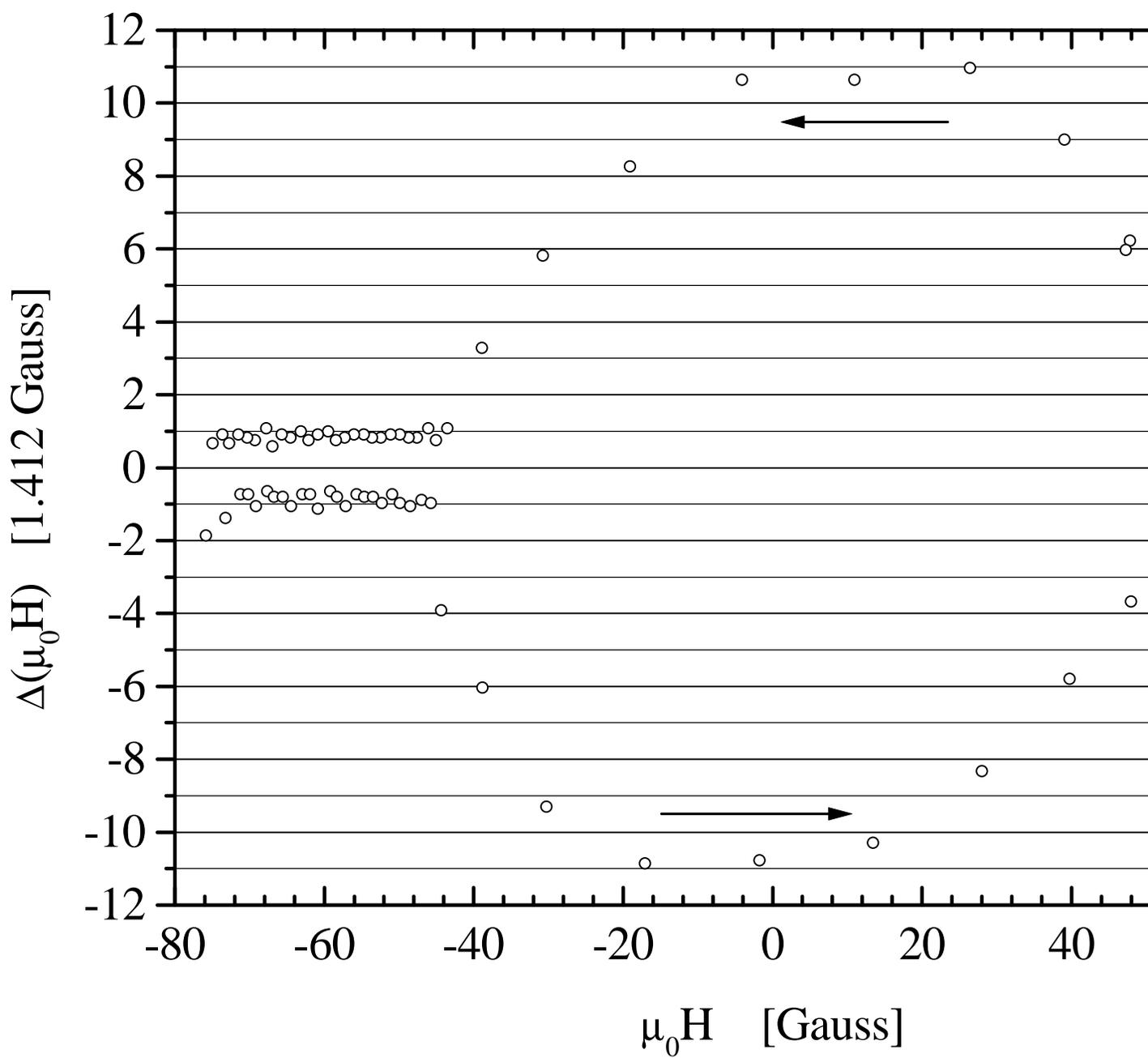
\end{document}